%% file: paper.tex
\documentclass[10pt, conference, letterpaper]{IEEEtran}
\IEEEoverridecommandlockouts

\usepackage{cite}
\usepackage{amsmath,amssymb,amsfonts}
\usepackage{graphicx}
\usepackage{textcomp}
\usepackage{xcolor}
\def\BibTeX{{\rm B\kern-.05em{\sc i\kern-.025em b}\kern-.08em
    T\kern-.1667em\lower.7ex\hbox{E}\kern-.125emX}}
\usepackage{comment}

\usepackage{graphicx}
\usepackage{wrapfig}
\usepackage{authblk}
\usepackage{textcomp}
\usepackage{xcolor}
\usepackage{wrapfig}
\usepackage{setspace}
\usepackage{color}
\usepackage[normalem]{ulem}
\usepackage{multirow}
\usepackage{hhline}
\usepackage{tikz}
\usepackage{url}
\usepackage{enumitem}
\usepackage{subcaption}
\usepackage{tikz}
\usepackage{kotex}

\usepackage{graphicx}
\usepackage{textcomp}
\usepackage{xcolor}
\usepackage{tcolorbox}
\usepackage[normalem]{ulem}
\usepackage{setspace}
\usepackage{colortbl}
\usepackage{xfp}

\usepackage[utf8]{inputenc}
\usepackage{kotex}
\usepackage{color}
\usepackage{algorithm}
\usepackage{algpseudocode}
\usepackage[normalem]{ulem}
\usepackage{makecell}
\usepackage{dblfloatfix} 
\usepackage{lipsum}
\usepackage{booktabs}
\usepackage{multirow}
\usepackage{siunitx}
\usepackage{verbatim}
\usepackage[colorlinks=true,linkcolor=violet,citecolor=violet,urlcolor=black]{hyperref}
\usepackage{circledsteps}

\newcommand{\CircledTextBlack}[2][]{%
    \CircledParamOpts{inner color=white, outer color=none, fill color=black, inner xsep=2pt, inner ysep=2pt, #1}{1}{#2}%
}

\usepackage{ulem}
\newcommand{\cb}{\textcolor{blue}}
\newcommand{\cred}{\textcolor{red}}

\newcommand{\squishlist}{
\begin{list}{$\bullet$}
	{ \setlength{\itemsep}{0pt}      \setlength{\parsep}{-0pt}
		\setlength{\topsep}{4pt}       \setlength{\partopsep}{0pt}
		\setlength{\listparindent}{-2pt}
		\setlength{\itemindent}{-5pt}
		\setlength{\leftmargin}{1em} \setlength{\labelwidth}{0em}
		\setlength{\labelsep}{0.5em} } }

\newcommand{\squishend}{
\end{list}}

\newcommand{\tbdfixed}{\textsc{{OpenCXD}}}

\begin{document}

%\title{Towards Accurate Evaluation of CXL-SSD Devices: Real-Device-Guided Simulator Design}
\title{\tbdfixed{}: An Open Real-Device-Guided Hybrid Evaluation Framework for CXL-SSDs}

%\author{Anonymous Author(s)}

%\begin{comment}
\author{
{Hyunsun Chung}$^{1,*}$, {\href{https://orcid.org/0009-0008-9293-173X}{Junhyeok Park}}$^{1,*}$, {Taewan Noh}$^{1}$, {Seonghoon Ahn}$^{1}$\\ 
%{Hyunsun Chung}$^{1}$, {Junhyeok Park}$^{1}$, {Taewan Noh}$^{1}$, {Seonghoon Ahn}$^{1}$\\ 
{Kihwan Kim}$^{1}$, 
{Ming Zhao}$^{2}$, 
%{Youngjae Kim}$^{1,\dagger{}}$
{Youngjae Kim}$^{1,\dagger{}}$
\thanks{$^{*}$They are first co-authors and have contributed equally.}
\thanks{$^{\dagger}$Y. Kim is the corresponding author.}
\\
{$^1$Sogang University, Seoul, Republic of Korea, 
$^2$Arizona State University, Tempe, AZ, USA}
}
%\end{comment}

\begin{comment}
\author{\IEEEauthorblockN{1\textsuperscript{st} Given Name Surname}
\IEEEauthorblockA{\textit{dept. name of organization (of Aff.)} \\
\textit{name of organization (of Aff.)}\\
City, Country \\
email address or ORCID}
\and
\IEEEauthorblockN{2\textsuperscript{nd} Given Name Surname}
\IEEEauthorblockA{\textit{dept. name of organization (of Aff.)} \\
\textit{name of organization (of Aff.)}\\
City, Country \\
email address or ORCID}
\and
\IEEEauthorblockN{3\textsuperscript{rd} Given Name Surname}
\IEEEauthorblockA{\textit{dept. name of organization (of Aff.)} \\
\textit{name of organization (of Aff.)}\\
City, Country \\
email address or ORCID}
\and
\IEEEauthorblockN{4\textsuperscript{th} Given Name Surname}
\IEEEauthorblockA{\textit{dept. name of organization (of Aff.)} \\
\textit{name of organization (of Aff.)}\\
City, Country \\
email address or ORCID}
\and
\IEEEauthorblockN{5\textsuperscript{th} Given Name Surname}
\IEEEauthorblockA{\textit{dept. name of organization (of Aff.)} \\
\textit{name of organization (of Aff.)}\\
City, Country \\
email address or ORCID}
\and
\IEEEauthorblockN{6\textsuperscript{th} Given Name Surname}
\IEEEauthorblockA{\textit{dept. name of organization (of Aff.)} \\
\textit{name of organization (of Aff.)}\\
City, Country \\
email address or ORCID}
}
\end{comment}

\maketitle

\setstretch{0.964}
\input{abs}

\begin{IEEEkeywords}   
Compute Express Link, Solid-State Drive
\end{IEEEkeywords}

\input{intro}

%\clearpage
\input{back}

\input{motiv}
\input{design}
\input{eval}

%\clearpage
\input{conc}

\section*{Acknowledgments}
{
This work was partially supported by the National Research Foundation of Korea (NRF) grants funded by the Korea  government (MSIT) (RS-2025-00564249 and RS-2024-00453929).
}

\begin{comment}
\section*{Acknowledgments}
\setstretch{0.53}{\scriptsize
We thank the anonymous reviewers of our paper for their invaluable comments on improving the paper.
This work was funded in part by the Institute of Information Communications Technology Planning Evaluation (IITP) grants funded by the Korea government (MSIT) (No. 2020-0-00104), 
in part by the National Research Foundation of Korea (NRF) grant funded by the Korean government (MSIT) (No. NRF-2021R1A2C2014386), and in part by SK hynix research grant.
This research used resources of the Oak Ridge Leadership Computing Facility at the Oak Ridge National Laboratory, which is supported by the Office of Science of the U.S. Department of Energy under Contract No. DE-AC05-00OR22725.
\iffalse This manuscript has been authored by UT-Battelle, LLC under Contract No. DE-AC05-00OR22725
with the U.S. Department of Energy. The publisher, by accepting the article for publication, acknowledges that the U.S.
Government retains a non-exclusive, paid up, irrevocable,
world-wide license to publish or reproduce the published
form of the manuscript, or allow others to do so, for United
States Government purposes. The Department of Energy will
provide public access to these results of federally sponsored
research in accordance with the DOE Public Access Plan
(http://energy.gov/downloads/doe-public-access-plan). \fi
}
\end{comment}

\scriptsize 
{%\setstretch{0.9}
  \bibliographystyle{ieeetr}
  \bibliography{paper}
}
\end{document}

%% file: abs.tex
\begin{abstract}
The advent of Compute Express Link (CXL) enables SSDs to participate in the memory hierarchy as large-capacity, byte-addressable memory devices. 
These CXL-enabled SSDs (CXL-SSDs) offer a promising new tier between DRAM and traditional storage, combining NAND flash density with memory-like access semantics. 
However, evaluating the performance of CXL-SSDs remains difficult due to the lack of hardware that natively supports the CXL.mem protocol on SSDs. 
As a result, most prior work relies on hybrid simulators combining CPU models augmented with CXL.mem semantics and SSD simulators that approximate internal flash behaviors. 
While effective for early-stage exploration, this approach cannot faithfully model firmware-level interactions and low-level storage dynamics critical to CXL-SSD performance.
In this paper, we present \tbdfixed{}, a real-device-guided hybrid evaluation framework that bridges the gap between simulation and hardware. 
\tbdfixed{} integrates a cycle-accurate CXL.mem simulator on the host side with a physical OpenSSD platform running real firmware. 
This enables in-situ firmware execution triggered by simulated memory requests.
Through these contributions, \tbdfixed{} reflects device-level phenomena unobservable in simulation-only setups, providing critical insights for future firmware design tailored to CXL-SSDs.
%\cred{In contrast to software-based simulation CXL-SSD evaluation platforms, where up to 94.3\% of NAND read latency values are the same, \tbdfixed{} reflects device-level phenomena unobservable in simulation-only setups, such as DRAM latency spikes over 2$\mu s$ and 2.4$\times$ higher NAND read latencies, providing critical insights for future firmware design tailored to CXL-SSDs.}

%\sout{These insights underscore the need for real-device feedback in designing and evaluating next-generation CXL memory systems.}
%\cred{yk: 마지막 문장 필요 없음}
\end{abstract}

%% file: intro.tex
\section{Introduction}
\label{sec:intro}

The growing scale of modern deep learning models and data analytics applications has led to memory footprints reaching tens of terabytes~\cite{kwon2018memorywall,derbyshire2025memory,zheng2023ocs}, far beyond the limits of traditional DRAM installations. 
This widening gap between demand and capacity has resurrected the infamous \textit{memory wall}~\cite{gholami2024memorywall}, in which memory bandwidth and size become critical bottlenecks to system performance. 
To mitigate this issue, researchers have begun exploring memory expansion techniques that repurpose alternative technologies as additional memory~\cite{oliveira2023extendingmemory, badam2011ssdalloc, jung2022cxlssd}, paving the way for new architectural paradigms. Among these, Compute Express Link (CXL)~\cite{cxl31spec} has emerged as a promising enabler of such memory expansion.
CXL is a high-bandwidth, cache-coherent interconnect built on top of the PCI Express (PCIe)~\cite{pcisig_specs} infrastructure. 
By using CXL, large-capacity PCIe devices like flash-based Solid-State Drives (SSDs) can be attached directly to the host system memory space, creating a memory pool that augments or disaggregates DRAM~\cite{das2024cxlintro}.

Notably, CXL’s memory semantics (CXL.mem) support byte-addressable access to device memory, meaning a CPU can read or write an SSD’s onboard DRAM buffer with ordinary load and store instructions. 
This eliminates the need for traditional I/O commands, significantly reducing software overhead and access latency while enabling a unified, tiered memory architecture that extends beyond DRAM’s capacity limits~\cite{zhang2025skybyte}. Building on this capability, a new class of memory-semantic devices, CXL-enabled SSDs (CXL-SSDs), has begun to emerge (§\ref{sec:back}). 
These devices leverage mature NAND flash technology to offer terabytes of byte-addressable capacity at a fraction of DRAM’s cost per gigabyte~\cite{jung2022cxlssd}, while maintaining access latencies in the microsecond range.
Although slower than DRAM by several orders of magnitude, CXL-SSDs provide significantly lower latency than traditional SSDs accessed via the block interface (e.g., NVMe~\cite{nvme_spec}). 

The key challenge for CXL-SSDs is how to architect and evaluate these devices effectively, maximizing their performance potential while addressing inherent latency trade-offs.
However, this evaluation remains difficult in practice, due to \textit{the lack of hardware that natively supports the CXL.mem protocol on SSDs}.
To overcome this, recent research has adopted software-based hybrid simulators, combining CPU simulators (e.g., MacSim~\cite{MacSim2025}, Gem5~\cite{gem5}) extended with CXL.mem semantics and SSD simulators (e.g., SimpleSSD~\cite{SimpleSSD2020}, FlashSim~\cite{kim2009flashsim}) that model internal flash behaviors such as address translation and I/O scheduling. 
This methodology enables early-stage design exploration and has been widely used in prior work~\cite{li2025bytefs,zhan2024romefs,wang2025cxlsim}.

Notably, the CXL.mem interface overhead itself has been characterized in prior studies~\cite{DasSharma2022FMS,MicrochipXpressConnect2020}, and shown to be relatively consistent and bounded~\cite{das2024cxlintro}.
As such, injecting this CXL interface time overhead into x86 simulations as a parameter is generally considered a reasonable approach for modeling host-side access costs.
However, \textit{the same cannot be said for modeling device-side behavior}.
Replacing a real SSD with a simulator introduces significant challenges in capturing the full complexity of CXL-SSD-specific firmware logic and storage interactions, which are limitations that fundamentally hinder accurate evaluation of CXL-SSD designs (§\ref{sec:motiv}).

There are two key limitations of simulation-only evaluation.
\noindent\textbf{First}, since CXL-SSDs function as memory rather than storage, they must handle fine-grained, cacheline-level memory accesses, making them highly sensitive to device-side performance fluctuations.
However, simulators typically rely on static latency models, overlooking dynamic behaviors such as real-time NAND latency variability and firmware delays, resulting in latency estimation errors as high as 36\%~\cite{simplessd_micro}.
\textbf{Second}, CXL-SSDs introduce new firmware-managed mechanisms, such as write logging and log compaction~\cite{zhang2025skybyte}, that do not exist in conventional SSDs. 
Simulating these new behaviors solely from the host side, without executing them on real hardware and capturing internal device interactions, fails to reflect crucial contention and scheduling effects, yielding an incomplete and often misleading performance picture~\cite{openssd_acm_paper}.

In this paper, we present \tbdfixed{}, a real-device-guided hybrid evaluation framework for CXL-SSDs. 
\tbdfixed{} combines a cycle-accurate CPU and memory simulator augmented with CXL.mem support and a hardware platform based on OpenSSD~\cite{openssd_acm_paper}, an open-source SSD prototype that runs actual controller firmware.
In essence, the host side of the system (CPU, cache, and memory controllers) is simulated, allowing full control over experimental scenarios and observability of system-level events, while the storage side is handled by a physical device that embodies a CXL-SSD (§\ref{sec:design}). 

We addresses two key challenges in developing \tbdfixed{}:

\squishlist
\item
\textbf{Enabling Cacheline-Level Access over NVMe.}
Unlike DRAM or DRAM-based CXL memory modules~\cite{SKhynixCXL2025}, publicly available SSD prototypes do not natively support cacheline-sized CXL.mem transactions.
Bridging this gap is non-trivial because NVMe operates on fixed-sized (e.g., 4~KB) blocks with DMA-based transfers, not on byte-addressable memory~\cite{bandslim, byteexpress}.
To address this, \tbdfixed{} defines custom NVMe commands that encode CXL.mem semantics and emulate cacheline-granularity memory access without modifying the physical hardware interface. 

\item
\textbf{Cycle-Level Timing Integration with Real Firmware.}
While the cycle-accurate host simulator tracks timing at the cycle level, the OpenSSD operates asynchronously with real-time firmware execution.
\tbdfixed{} resolves this by adopting a \textit{device-in-the-loop} design: for each memory access, the simulator pauses execution, delegates the operation to OpenSSD running real firmware, and waits until the firmware measures the end-to-end latency and reports it to the host simulator.
The simulator then resumes by converting the measured latency to cycles using a calibrated timing ratio and advances its internal clock accordingly.
\squishend

We then implement key firmware components of a state-of-the-art CXL-SSD~\cite{zhang2025skybyte} within the OpenSSD device. 
This allows fine-grained memory requests from the host to trigger CXL-SSD firmware execution.
These design choices enable \tbdfixed{} to maintain the configurability and observability of full-system simulation, while introducing \textit{hardware realism} through in-situ execution of the SSD’s software stack.

Our evaluation shows that \tbdfixed{} captures performance characteristics from real-life hardware unobservable in software SSD simulations. These include 2.4$\times$ higher NAND read latencies due to lower-level NAND controller and SSD firmware overheads, as well as DRAM latency spikes over 2$\mu s$.
Such evaluation highlight the need for real-device-guided evaluation to uncover nuanced CXL-SSD dynamics such as out-of-order persistence and timing variability.

This work makes the following contributions:

\squishlist
\item \textbf{Real-Device-Guided CXL-SSD Evaluation Framework}: We present \tbdfixed{}, the first hybrid platform that combines a full-system CXL.mem simulator with an open-source SSD prototype running real firmware, enabling accurate and practical evaluation of CXL-SSD architectures.

\item \textbf{Device-in-the-Loop Architecture}: We develop a tightly coupled host-device interface that allows simulated memory accesses to trigger firmware execution and measure and integrate its latency into host-side simulation in situ.

\item \textbf{Implementation of CXL-SSD Internals on Hardware}: We implement key CXL-SSD components, including write log and log index, on actual hardware, revealing behaviors and bottlenecks that prior simulation-only setups miss.
\squishend

%% file: back.tex
\section{Background}
\label{sec:back}

\subsection{Hardware Architecture of CXL-SSDs} 
\label{sec:back_hw_arch}

\begin{figure}[!t]
%    \vspace{-6pt}
    \centering	\includegraphics[width=0.85\linewidth]{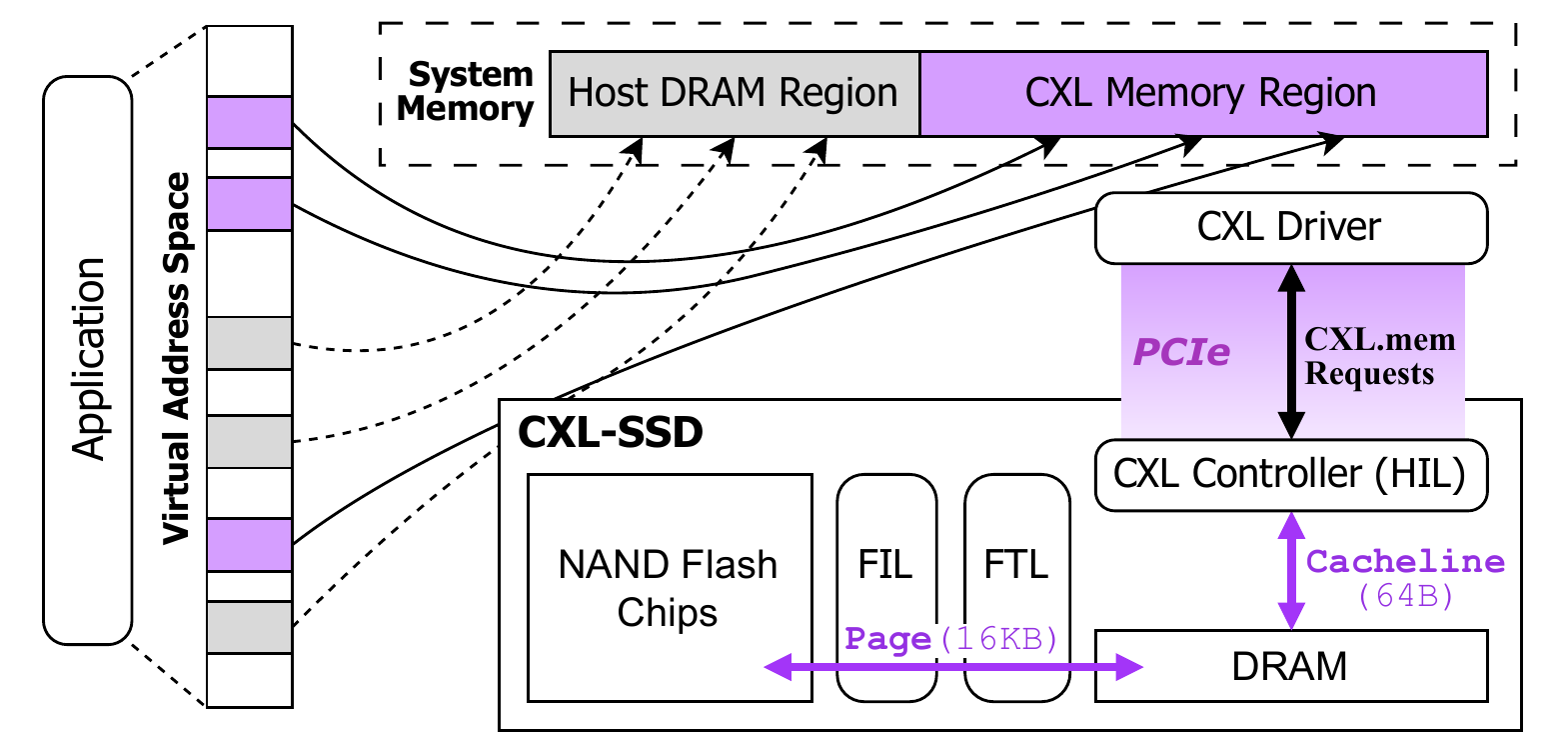}
    \vspace{-3pt}
    \caption{System architecture of CXL-SSD.}
    \vspace{-10pt}
    \label{fig:back_cxlssd}
\end{figure}

The hardware architecture of CXL-SSDs is composed of the same core components as conventional SSDs~\cite{jung2022cxlssd,kwon2023cxlprefetch,zhang2025skybyte}, including NAND, DRAM, and a System-on-Chip (SoC) controller.
CXL-SSDs differ from traditional SSDs in how the device is integrated into the system: they function as CXL.mem endpoints, exposing their capacity as part of the host’s memory address space.
When the CPU accesses a CXL memory region, cacheline-sized memory requests (typically 64~B) are issued via the CXL driver, and the Host Interface Layer (HIL) of the SSD controller buffer them in the on-board DRAM, and eventually flush them to NAND in page-sized units (e.g., 16~KB) through the Flash Translation Layer (FTL) and Flash Interface Layer (FIL), as illustrated in Fig.~\ref{fig:back_cxlssd}.

\renewcommand{\thesubfigure}{\textbf{\alph{subfigure}}}

\subsection{Software Architecture of CXL-SSDs} 
\label{sec:back_sota}

\begin{figure}[!t]
    \centering
    \begin{subfigure}{0.93\linewidth}
        \centering
        \includegraphics[width=\linewidth]{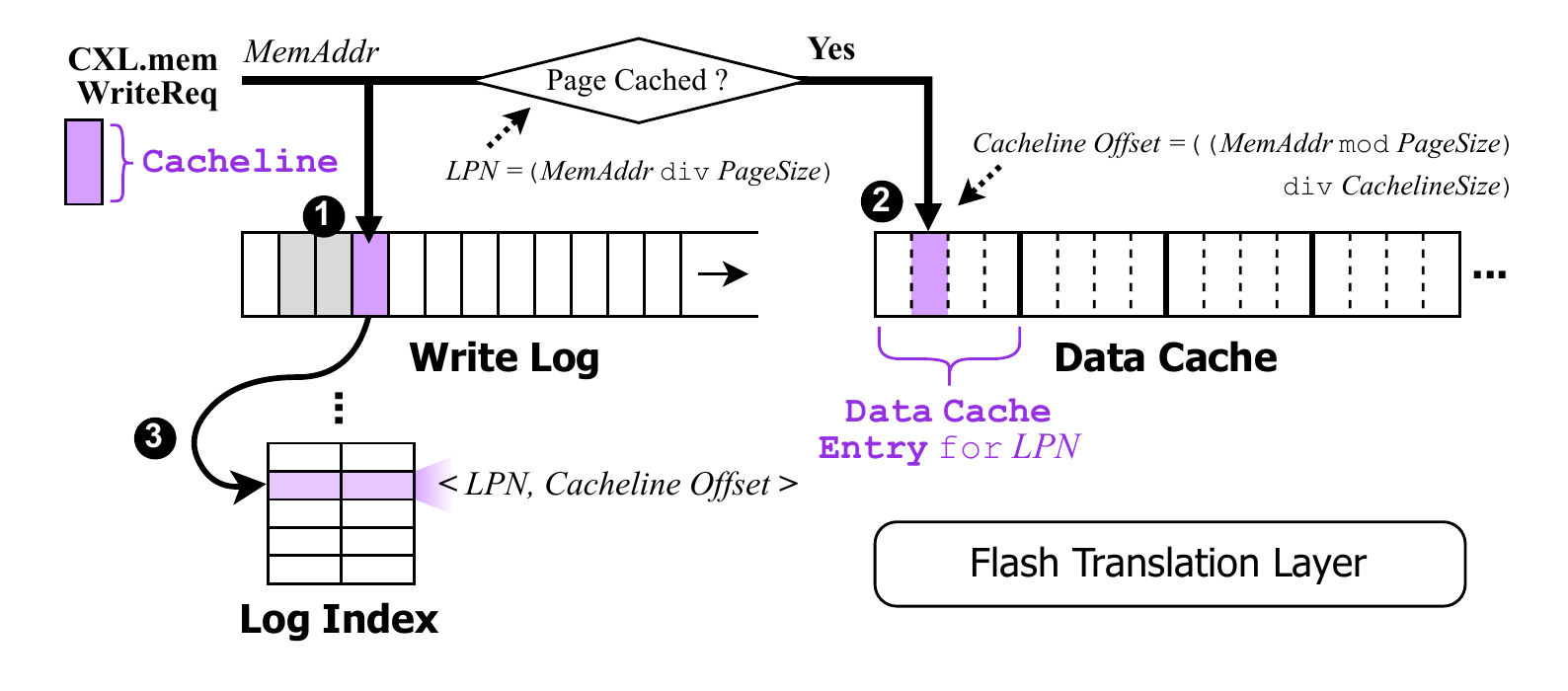}
        \vspace{-25pt}
        \caption{\textbf{Write operation}}
        \vspace{5pt}
    \end{subfigure}
    
    \begin{subfigure}{0.93\linewidth}
        \centering
        \includegraphics[width=\linewidth]{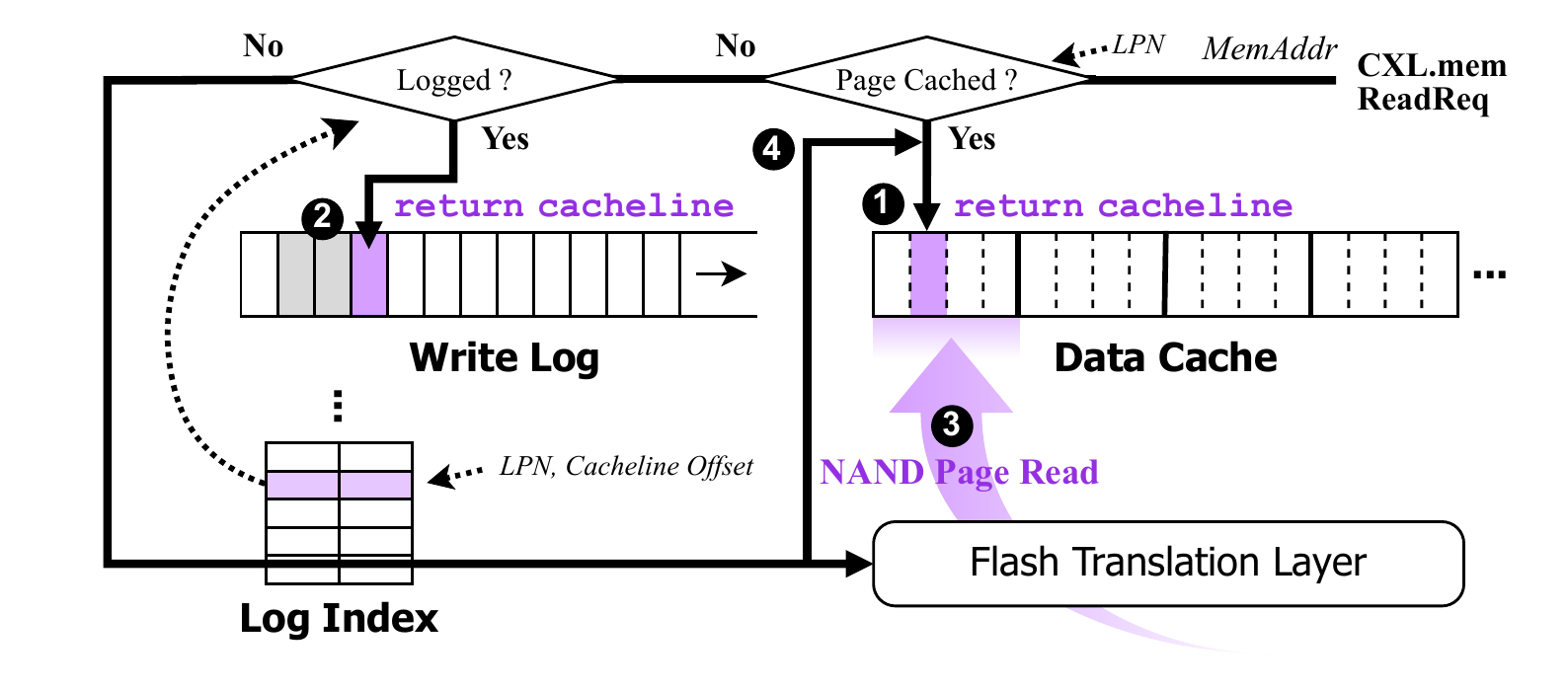}
        \vspace{-25pt}
        \caption{\textbf{Read operation}}
    \end{subfigure}
    
    \vspace{-3pt}
    \caption{Write/read flows of the state-of-the-art CXL-SSD~\cite{zhang2025skybyte}, comprising a \textit{Write Log}, \textit{Data Cache}, and \textit{Log Index}.}
    \vspace{-10pt}
    \label{fig:back_sota}
\end{figure}

SkyByte~\cite{zhang2025skybyte} represents the state-of-the-art design of a CXL-SSD and provides a detailed explanation of its internal software architecture, as illustrated in Fig.~\ref{fig:back_sota}.
The key software components inside the CXL-SSD include a \textit{Write Log}, \textit{Data Cache}, and \textit{Log Index}, which bridge the granularity mismatch between 64~B cacheline accesses and NAND page-level operations.
The \textit{Write Log} buffers incoming CXL.mem write requests at 64~B granularity. 
The \textit{Data Cache} functions similarly to a traditional SSD’s in-memory NAND page cache, managing recently accessed NAND pages.
Additionally, to mitigate the latency penalty from NAND I/O, SkyByte incorporates context switching to perform other I/O requests while said NAND I/O operation is ongoing.

Fig.~\ref{fig:back_sota}(a) illustrates the write path.
When a CXL.mem write arrives, \CircledTextBlack{1} the device first stores the cacheline-sized payload into the \textit{Write Log}. 
\CircledTextBlack{2} If the corresponding NAND page is resident in the \textit{Data Cache}, the cacheline update is also applied there to maintain consistency. 
\CircledTextBlack{3} The system then updates the \textit{Log Index} to track the location of the buffered write.

Fig.~\ref{fig:back_sota}(b) illustrates the read path.
When a CXL.mem read arrives, \CircledTextBlack{1} the system checks whether the target cacheline is present in the \textit{Data Cache}; if so, it is returned directly to the host. 
\CircledTextBlack{2} If the \textit{Data Cache} does not contain the cacheline but the \textit{Write Log} does, the system retrieves and returns the buffered version. 
In cases where the cacheline is found in neither buffer, \CircledTextBlack{3} the corresponding NAND page is read into the \textit{Data Cache}, and \CircledTextBlack{4} the cacheline is served from there.
During this page load, existing NAND pages in the \textit{Data Cache} may be evicted and flushed back to NAND.

To manage and track buffered writes, SkyByte employs a two-level \textit{Log Index}: the first level identifies modified NAND pages, while the second level maps individual cacheline offsets within those pages. 
To reclaim log space and persist updates, the system periodically performs \textbf{log compaction}.
During compaction, the system first scans the first-level index to identify NAND pages that contain valid log entries. 
If such a page is already present in the \textit{Data Cache}, it is flushed directly to NAND. 
Otherwise, the NAND page is loaded into memory. 
It then consults the second-level index to locate all valid, buffered cachelines associated with the page and merges them into the in-memory copy. 
Finally, the merged page is written back to NAND, and the log entries are invalidated.

%% file: motiv.tex
\section{Motivation}
\label{sec:motiv}

\subsection{Limitations of Purely Software-Driven SSD Simulation}

\noindent\textbf{Limitation \#1) Reliance on Parameter-Driven Calculation:} 
Accurate characterization of NAND I/O latency is critical for optimizing CXL-SSDs, where all memory accesses operate at cacheline granularity and DRAM cache misses directly translate to NAND operations (§\ref{sec:back_sota}). As CXL-SSDs function as memory-semantic devices, they are highly sensitive to the latency path between on-board DRAM and NAND.
Therefore, understanding and optimizing these paths require precise visibility into device-level behavior.

However, most SSD simulators, including SimpleSSD~\cite{SimpleSSD2020} used in SkyByte, rely on parameter-driven, static latency modeling to estimate NAND flash behavior~\cite{simplessd_pal_2024}.
These models use predefined latency values of the NAND flash used in the SSD as input parameters to derive NAND read and program latency~\cite{simplessd_pal2_cc}, ignoring dynamic factors such as flash controller scheduling and internal firmware optimizations.
While efforts have been made to address this issue, for example, SimpleSSD now models ARM cores and flash parallelism (channels, ways, dies, planes) to simulate NAND I/O scheduling, these remain abstractions built atop assumed hardware parameters. 
Indeed, SimpleSSD reports deviations of up to 28\% in throughput and 36\% in latency compared to real hardware~\cite{simplessd_micro}.
This limitation is also evident in our experiments (§\ref{sec:motiv_nand_io}), where static latency modeling fails to reflect real-world NAND I/O behavior.

\noindent\textbf{Limitation \#2) Missing Hardware-Grounded Firmware Validation:}
A side effect of using software-driven SSD simulations is as no real SSD hardware is involved, all code related to SSD firmware is run on a host system.
This environment can lead to inaccurate assumptions of what is possible in SSD firmware for implementing CXL-SSD optimizations. 
SkyByte works around this limitation by implementing their optimizations, such as the \textit{Write Log}, \textit{Log Index}, and \textit{Data Cache}, in prototype FPGA hardware, and use the averages of operation measurements in the simulation~\cite{zhang2025skybyte}.
While this approach enables partial evaluation grounded in real hardware, it comes short of a holistic assessment across the full CXL-SSD stack. 
Specifically, it cannot capture the dynamic interaction among DRAM, NAND flash, and ARM cores triggered by real-time CXL.mem requests during workload execution (§\ref{sec:eval}).

The approach also faces previously mentioned pitfalls of simulations using statically declared parameters for operational overheads in CXL-SSD evaluation.  
Although certain firmware components are executed on prototype hardware, the measured latencies are averaged and reused as static inputs in the simulator, making the evaluation effectively parameter-driven.  
As a result, it cannot capture the dynamic, request-level interactions that emerge from executing the full firmware stack on real SSD hardware under live CXL.mem workloads.

\vspace{-10pt}
\begin{figure}[b!]
    \begin{subfigure}{\linewidth}
    \centering
    \includegraphics[width=0.9\linewidth]{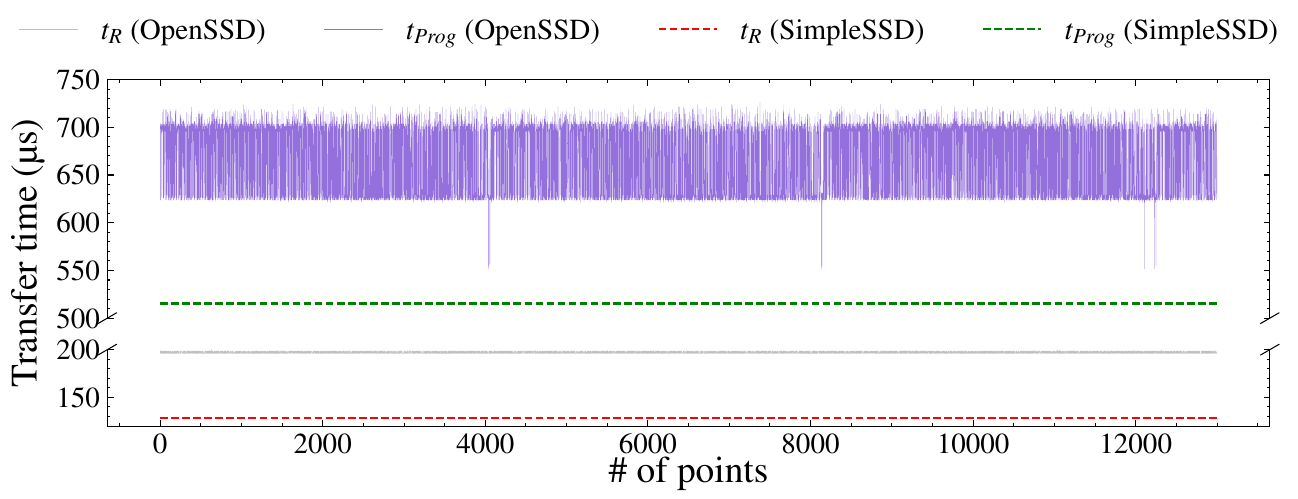}
    \vspace{-5pt}
    \caption{}
    \label{subfig:fio_toshiba_1}
    \end{subfigure}
    \begin{subfigure}{\linewidth}
    \centering
    \includegraphics[width=0.9\linewidth]{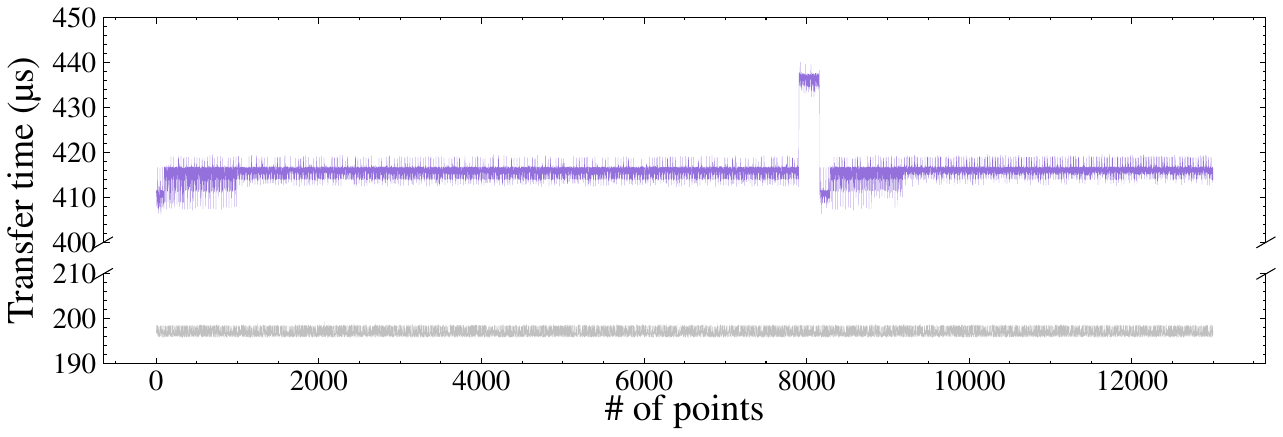}
    \vspace{-5pt}
    \caption{}
    \label{subfig:fio_sk_1}
    \end{subfigure}
    \vspace{-5pt}
    \caption{NAND read/program I/O times of two different types of NAND \textbf{(a)} and \textbf{(b)} with \texttt{iodepth=1}.}\label{fig:realnand-1}
%    \vspace{-10pt}
\end{figure}

\begin{figure}[h!]
    \begin{subfigure}{\linewidth}
    \centering
    \includegraphics[width=0.9\linewidth]{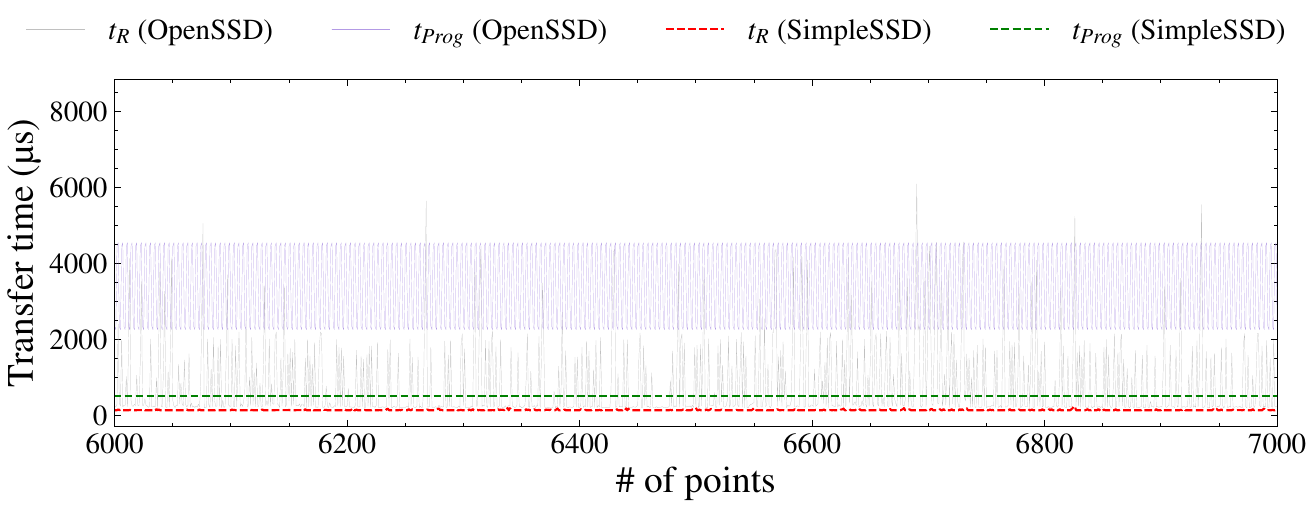}
    \vspace{-5pt}
    \caption{}
    \label{subfig:fio_toshiba_8}
    \end{subfigure}
    \begin{subfigure}{\linewidth}
    \centering
    \includegraphics[width=0.9\linewidth]{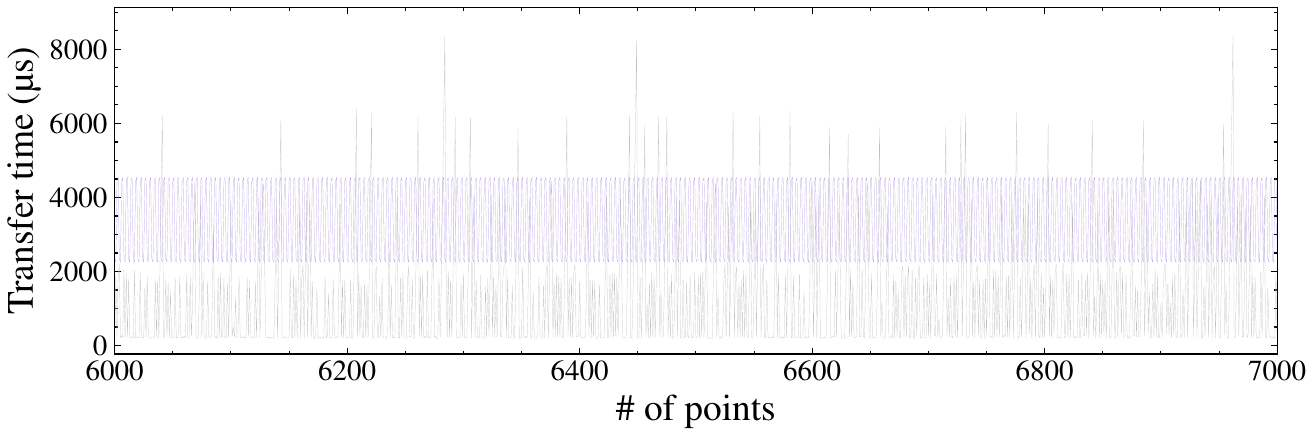}
    \vspace{-5pt}
    \caption{}
    \label{subfig:fio_sk_8}
    \end{subfigure}
    \vspace{-5pt}
    \caption{NAND read/program I/O times of two different types of NAND \textbf{(a)} and \textbf{(b)} with \texttt{iodepth=8}. The data is zoomed in to show the 6000-7000 range for clarity.}\label{fig:realnand-8}
    \vspace{-10pt}
\end{figure}

\subsection{Real-Life NAND I/O Characteristics in SSDs}
\label{sec:motiv_nand_io}

To show the degree of deviation of real-life hardware from a theoretical simulation, we measured NAND read/program I/O request times of two different NAND modules from \textbf{(a)} SK Hynix and \textbf{(b)} Toshiba using the OpenSSD platform~\cite{openssd_acm_paper}. The specification of said NAND modules can be seen in Table~\ref{tab:nand_specs}.
The measurements were taken using the \texttt{randread} and \texttt{randwrite} benchmarks of \texttt{fio}~\cite{AxboeFio2022} with an I/O unit of the NAND page size. The specifications of the OpenSSD platform and host workstation are in Tables~\ref{tab:eval_setup_openssd} and \ref{tab:eval_setup_host}.

\begin{table}[h!]
\footnotesize
\caption{Specifications of NAND flash modules used.}
\vspace{-5pt}
\centering
\begin{tabular}{|c||c|c|c|} 
\hline
\cellcolor{gray!45}Manufacturer & \cellcolor{gray!15}Capacity   & \begin{tabular}[c]{@{}c@{}}\cellcolor{gray!15}Parallelism\\\cellcolor{gray!15}Setup\end{tabular}                 & \begin{tabular}[c]{@{}c@{}}\cellcolor{gray!15}NAND Flash\\\cellcolor{gray!15}Page Size\end{tabular}  \\ 
\hline
\cellcolor{gray!15}\textbf{(a)} SK Hynix     & 1~TiB   & \multirow{2}{*}{\begin{tabular}[c]{@{}c@{}}4 Channel,\\8 Way\end{tabular}} & \multirow{2}{*}{16~KiB}                              \\ 
\hhline{|-||-|~~|}
\cellcolor{gray!15}\textbf{(b)} Toshiba      & 256~GiB &                                                                            &                                                     \\
\hline
\end{tabular}
\vspace{-0pt}
\label{tab:nand_specs}
\end{table}

Measurements of the time between when the NAND I/O request is issued to the low-level NAND flash controller by the SSD firmware, and when the firmware receives confirmation of request completion by the NAND flash controller were taken.
For comparison with the SSD simulation platform, results from SimpleSSD was measured using its built-in I/O request generating benchmark, with simulation parameters taken from NAND \textbf{(a)} specifications.   
As SimpleSSD experiments used only NAND \textbf{(a)} parameters, the results are shown exclusively in figures based on that setup.

Fig.~\ref{fig:realnand-1} shows the results with one outstanding I/O request, and reveals the differing performance characteristics of real and simulated NAND. 
Comparing NAND \textbf{(a)}'s read and program latency from OpenSSD and SimpleSSD shows that the real life NAND latency measurements are higher than the simulated results. 
Such results were observed as the main contributor to NAND latency is the provided NAND read/program latency parameters in the simulator.

With a workload with higher outstanding I/O requests, deviation from the simulation results exacerbates, as seen in Fig.~\ref{fig:realnand-8} which shows the results with eight outstanding I/O requests. Table \ref{tab:stdev_motiv} shows the disparity of real and simulated NAND latency variance, with simulated NAND latency staying close to the given latency parameters, while real NAND latency's standard deviation grows to the thousand of microseconds. While SimpleSSD does show small amounts of deviation for $t_{R}$ due to its timeline scheduling of NAND I/O, it is not able to cover the extent of actual deviation. 

\begin{table}
\centering
\caption{Standard deviation of NAND read ($t_{R}$) and program ($t_{Prog}$) times of NAND \textbf{(a)}, \textbf{(b)}, and SimpleSSD's simulation, with \texttt{iodepth=1} and  \texttt{iodepth=8}. Units are in microseconds.}
\footnotesize
\begin{tabular}{ccclcc} 
\toprule
\multirow{2}{*}{\begin{tabular}[c]{@{}c@{}}NAND\\Type\end{tabular}} & \multicolumn{2}{c}{iodepth=1} & \multicolumn{1}{c}{} & \multicolumn{2}{c}{iodepth=8}  \\ 
\cmidrule(l){2-3}\cmidrule(l){5-6}
                                                                    & $t_R$ & $t_{Prog}$              &                      & $t_R$   & $t_{Prog}$             \\
\textbf{(a)}                                                            & 1.1   & 37.61                 &                      & 974.16  & 1110.91              \\
\textbf{(b)}                                                             & 0.89  & 3.19                  &                      & 1374.84 & 1107.97              \\
\multicolumn{1}{l}{SimpleSSD}                                       & 0     & 0                     & \multicolumn{1}{c}{} & 11.1    & 0                    \\
\bottomrule
\end{tabular}
\vspace{-4pt}
\label{tab:stdev_motiv}
\end{table} 

Such a difference in latency between real and simulated NAND I/O is due to the low-level NAND flash controller and SSD firmware overhead not being taken into consideration in the simulations. 
These additional factors that add additional latency therefore act as blind spots in the simulation accuracy.

\begin{figure}[!h]
    \centering	\includegraphics[width=0.95\linewidth]{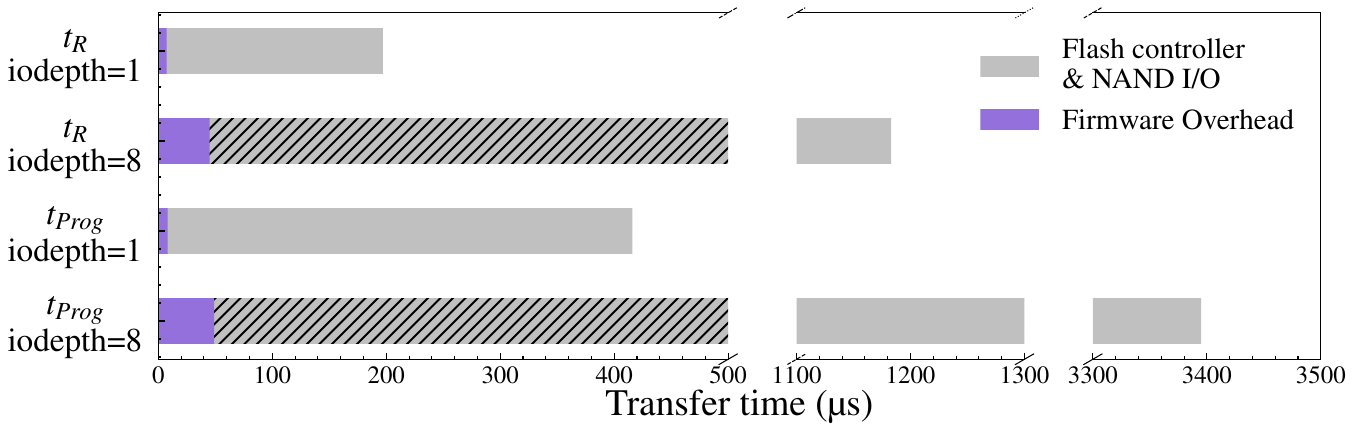}
    \vspace{-3pt}
    \caption{Breakdown of NAND \textbf{(b)}'s average $t_{R}$ and $t_{Prog}$.}
    \vspace{-0pt}
    \label{fig:motiv_breakdown}
\end{figure}

Fig.~\ref{fig:motiv_breakdown} demonstrates the significance of these blind spots by breaking down the average NAND I/O time of NAND \textbf{(b)}. The breakdown shows the presence of flash controller and firmware overhead, as well as a positive correlation with outstanding I/O requests and NAND I/O latency. This trend is further verified by the latency values documented by OpenSSD paper~\cite{openssd_acm_paper} showing the same trends as seen in our experimental results.

\vspace{0pt}
\begin{figure}[h]
\captionsetup[subfigure]{justification=centering}
\centering
\subfloat[]{\label{fig:cdf-a} \includegraphics[width=0.163\textwidth]{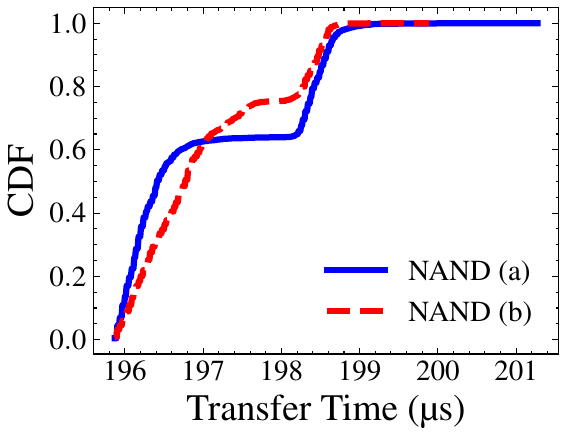}\vspace{-5pt}}% The "%" masks the line break.
\hfill
\subfloat[]{\label{fig:cdf-b} \includegraphics[width=0.152\textwidth]{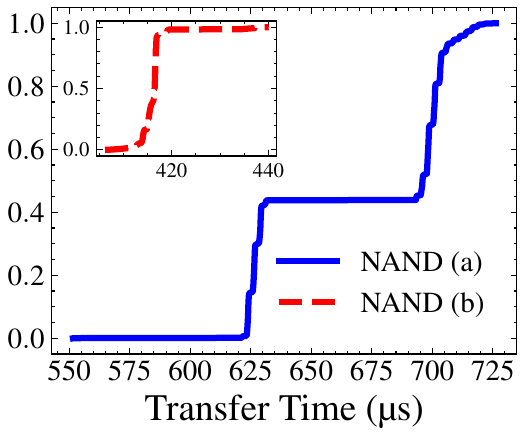}\vspace{-5pt}}%
\hfill
\subfloat[]{\label{fig:cdf-c} \includegraphics[width=0.152\textwidth]{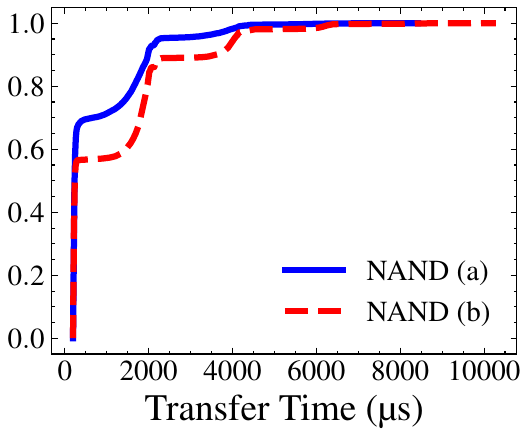}\vspace{-5pt}}%
\hfill
\vspace{-5pt}
\caption{NAND I/O latency Cumulative Distribution Function (CDF) of two different types of NAND in different workloads (a) \texttt{randread}, \texttt{iodepth=1}, (b) \texttt{randwrite}, \texttt{iodepth=1}, (c) \texttt{randread}, \texttt{iodepth=8}.} 
\label{fig:latency_cdf}
\vspace{-0pt}
\end{figure}

A deeper analysis of the real NAND results reveals the two NAND types also show different latency characteristics between each other.
To demonstrate, Fig.~\ref{fig:latency_cdf} shows NAND I/O latency CDF graphs of the three previous \texttt{fio} benchmarks. Even with benchmarks where both NAND types show overlapping ranges of latency values, each NAND module shows differing distributions of latency for all benchmarks.
Plus, returning to Fig.~\ref{fig:realnand-1}, the NAND \textbf{(b)} interestingly shows a brief spike in latency of up to 440$\mu s$ before returning to near-median values, an effect that cannot be captured by simulators.

\noindent\textbf{Bridging Simulation and Reality:}
Our findings suggest that while NAND flash latency specifications are useful, \textit{they do not fully capture the nuances of per-request performance in real-world scenarios}.
In other words, although SSD simulators are effective for estimating average performance and extracting architectural insights, they often diverge from real behavior when used to evaluate fine-grained, real-time performance at the request level.
This gap becomes especially critical in the context of CXL-SSDs, where the device functions more like memory than traditional block storage.

To address this issue, we explore the integration of the OpenSSD hardware platform with a host-side x86 simulator to construct a hybrid evaluation environment for CXL-SSDs.
By \textit{\textbf{combining the cycle-level accuracy of x86 simulation with the realism of actual SSD hardware execution}}, this approach enables the development and evaluation of CXL-SSD internal software components with both practical feasibility and performance characteristics that closely reflect real behavior.
Building on this foundation, we propose \textbf{\tbdfixed{}}, the first real-device-guided CXL-SSD simulation platform.

%% file: design.tex
\section{\tbdfixed{}: Proposed System}
\label{sec:design}

\begin{figure}[!t]
    \centering	
    \vspace{-7pt}
    \includegraphics[width=0.68\linewidth]{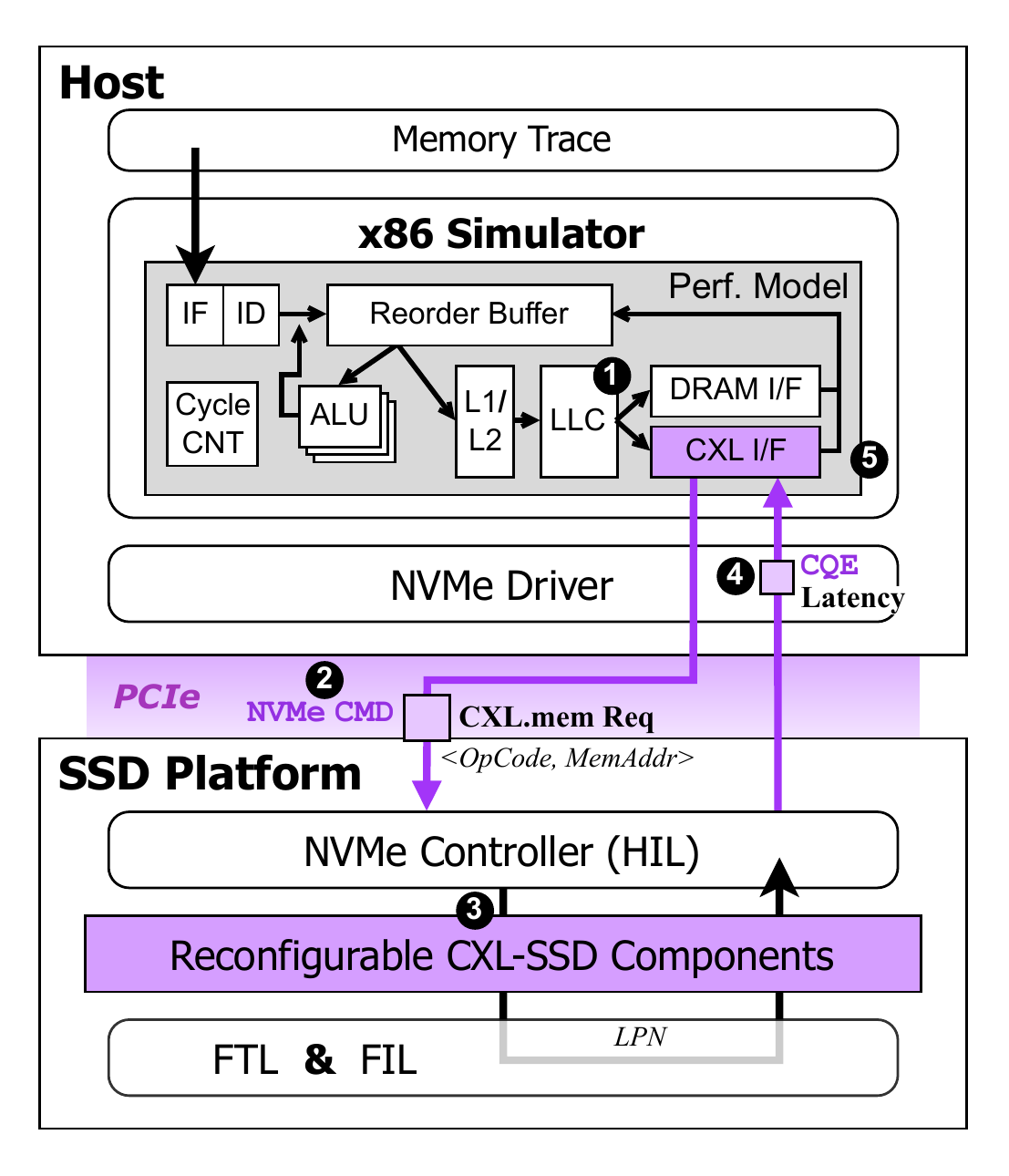}
    \vspace{-7pt}
    \caption{Architecture of \tbdfixed{} and its execution flow.}
    \vspace{-10pt}
    \label{fig:design_overview}
\end{figure}

\subsection{Architectural Overview and Operation Flow}

The overall architecture consists of two main components: \textit{x86 simulator} and \textit{SSD platform}.
The \textbf{\textit{x86 simulator}} provides cycle-accurate simulation of the entire memory hierarchy and replays memory traces derived from the target workload to reconstruct its instruction execution flow. 
It models detailed behaviors of the multi-layered cache hierarchy and the memory interface, allowing simulation of scenarios where Last-Level Cache (LLC) misses trigger memory accesses to either the main system memory or CXL memory.
\tbdfixed{} leverages this capability by intercepting LLC misses and, when the target address falls within a CXL-mapped range, redirecting the memory request to the SSD platform.
The \textbf{\textit{SSD platform}} is built on an OpenSSD platform~\cite{openssd_acm_paper}, which replicates the hardware architecture of a commercial SSD, incorporating a SoC controller, DRAM, and NAND flash chips. 
The SoC runs a SSD firmware, including the HIL, FTL, and FIL (§\ref{sec:back_hw_arch}). 

As illustrated in Fig.~\ref{fig:design_overview}, the execution flow of \tbdfixed{} proceeds as follows:
\CircledTextBlack{1} When an LLC miss occurs, the x86 simulator checks whether the missed address falls within the memory-mapped region assigned to the CXL-SSD.
\CircledTextBlack{2} If it does, the request is encapsulated into a custom NVMe command that explicitly encodes the CXL.mem access semantics for \tbdfixed{}, and issued to the SSD platform via NVMe passthrough~\cite{libnvme_ioctl}.
\CircledTextBlack{3} On the SSD side, the NVMe controller fetches the command and performs the corresponding CXL-SSD operation (e.g., write logging) based on the embedded request information. During this time, the x86 simulator pauses its cycle count.
\CircledTextBlack{4} The total device latency is measured on the SSD and returned to the host by embedding it in a reserved field of the Completion Queue Entry (CQE)~\cite{nvme_spec}.
\CircledTextBlack{5} The NVMe driver extracts this latency from the CQE and reports it to the x86 simulator, which adds the CXL.mem interface overhead and integrates the total latency into its cycle count with resuming its cylce count progression.

\subsection{x86 Simulator with CXL-SSD Integration}
\label{sec:host_simul}

\begin{figure}[!h]
    \centering	
    \vspace{-7pt}
    \includegraphics[width=0.95\linewidth]{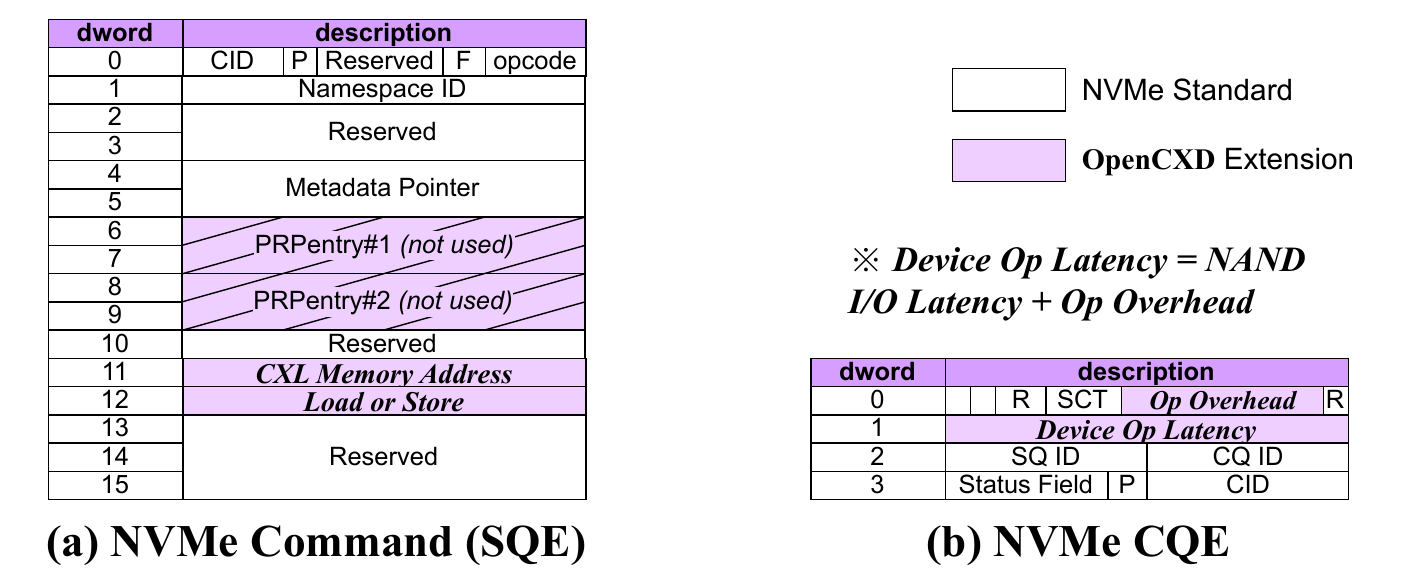}
    \vspace{-3pt}
    \caption{NVMe command and CQE for \tbdfixed{}.}
    \vspace{-0pt}
    \label{fig:design_nvme_cmd}
\end{figure}

\noindent\textbf{CXL Memory Request over NVMe:}
We extend the memory subsystem of the x86 simulator with a custom CXL.mem path that integrates with the SSD platform. 
This integration allows memory requests targeting the CXL-SSD region to be redirected as NVMe commands to the device.
Specifically, the x86 simulator distinguishes memory requests by inspecting the target address of each 64-byte cacheline access. 
If the address falls within a CXL-SSD-mapped region, the simulator constructs a custom NVMe command that embeds the memory address and opcode of the memory request (see Figure~\ref{fig:design_nvme_cmd}(a)). 
Note that NVMe's data payload transfer is intentionally disabled~\cite{bandslim}, as the x86 simulators, including MacSim~\cite{MacSim2025}, typically do not model and process actual data payloads.

\noindent\textbf{Device-in-the-Loop Timing Integration:}
Once the command is constructed, the x86 simulator issues it to the SSD platform via NVMe passthrough~\cite{libnvme_ioctl}, then pauses its cycle progression and waits.
Upon receiving the command, the SSD controller starts measuring the latency of the CXL-SSD operation.
When the operation completes, the controller stops the timer, embeds the measured latency into a reserved field of the CQE, and returns the CQE to the host.
To support fine-grained performance analysis and context switches used in CXL-SSD optimizations, the controller also extracts and reports CXL operation overhead separately from the total device latency (see Figure~\ref{fig:design_nvme_cmd}(b)).
Then, the simulator resumes execution by adding the CXL.mem interface overhead to the device-measured latency and converting the total delay into cycles based on its internal clock frequency.

% {
% \vspace{-5pt}
% \footnotesize
% \begin{equation}
% \label{eq:how_to_integrate}
% \mathit{DevLatency} \mathrel{+}= \mathit{CxlMemCost}, \quad 
% \mathit{Tick}_{\text{cur}} \mathrel{+}= \frac{\mathit{DevLatency}}{\mathit{ClockFreq}}
% \end{equation}
% }

To model the CXL.mem interface delay, \tbdfixed{} applies a configurable overhead value.
As prior work has shown~\cite{das2024cxlintro}, the CXL.mem interface delay is generally stable; for instance, SkyByte~\cite{zhang2025skybyte} assumes a fixed latency of 40~ns, which we also adopt in our evaluation.
This delay parameter in \tbdfixed{} can be adjusted to emulate different CXL interface designs.

\begin{figure}[!h]
    \centering	
    \vspace{-10pt}
    \includegraphics[width=1.0\linewidth]{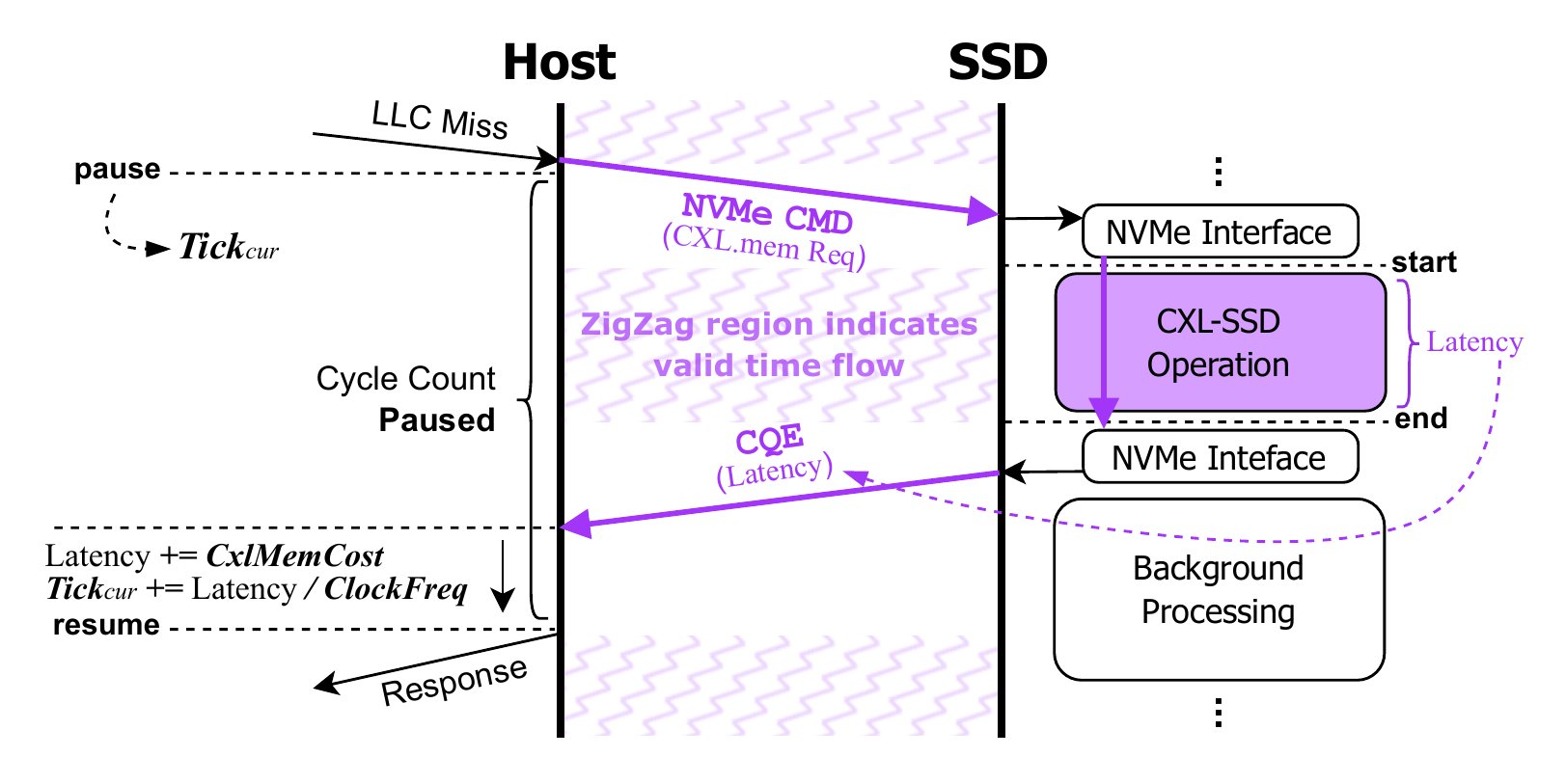}
    \vspace{-15pt}
    \caption{Timing flow of a CXL memory request in \tbdfixed{}.}
    \vspace{-0pt}
    \label{fig:design_timing}
\end{figure}

As shown in Fig.~\ref{fig:design_timing}, \tbdfixed{} enables accurate \textit{device-in-the-loop integration} by suspending the x86 simulator during each CXL-SSD access and incorporating measured firmware-level latencies.
This approach faithfully models CXL-SSD timing behavior while excluding NVMe communication overhead, which would not be present in real CXL-SSDs.

\subsection{SSD Platform with CXL-SSD Firmware Support}

To reproduce the behavior of CXL-SSDs, we implement key software components described in Section~\ref{sec:back_sota} on the SSD platform.
These include a \textit{Write Log}, a \textit{Data Cache}, and a \textit{Log Index} with log compaction support.
In a conventional SSD, incoming NVMe commands are typically handled by the HIL using a block I/O handler.
In contrast, \tbdfixed{} defines a new handler tailored for CXL-SSD operations.
This custom handler first extracts the memory address and opcode (e.g., Read, Write) from the given CXL.mem memory request.
It then performs the corresponding operation based on the opcode.
For example, in the case of a write, the handler appends the data to the current index of the \textit{Write Log}.
If the \textit{Data Cache} already contains the NAND page that includes the target cacheline, the handler updates the relevant cacheline offset within that cached page.
Then, the \textit{Log Index} is updated to record the cacheline’s location, as illustrated in Fig.~\ref{fig:back_sota}.

%\subsection{Concurrent Access to the CXL-SSD}

\subsection{Concurrent Access Modeling and Future Extensions}

\tbdfixed{} in its current form can simulate concurrent CXL-SSD access on the x86 simulator level. By replaying workload traces collected from 8 Skylake cores with 3 threads each---naturally containing interleaved CXL.mem accesses---\tbdfixed{} reflects much of the timing behavior of concurrent requests from a host system.
%\cb{However, a limitation exists in completely modeling true in-device parallel execution. As the x86 simulator issues memory requests to the SSD platform with NVMe passthrough via ioctl, each memory request is processed sequentially within the SSD platform. Such a limitation can result in less accurate representation of multiple consecutive memory requests that all result in cache misses and perform concurrent NAND I/O.}
%\cb{Addressing this limitation is planned for future work, as it will enable \tbdfixed{} to better represent performance characteristics of highly parallel CXL-SSD architectures.}
While the current design effectively models host-level concurrency, it focuses on sequential request processing within the SSD platform when interfacing through NVMe passthrough via ioctl. 
This design choice simplifies the integration between the x86 simulator and the SSD platform, ensuring consistent timing control and reproducibility during trace replay. 
However, it also means that simultaneous in-device processing paths, such as overlapping NAND I/O operations or internal command pipelining, are not exercised. This limitation may lead to an underestimation of performance in workloads with a high consecutive cache miss ratio in the CXL-SSD. Enhancing this aspect is planned for future work, which will enable \tbdfixed{} to more extensively reflect performance characteristics of highly parallel CXL-SSD architectures.

%% file: eval.tex
\section{Evaluation}
\label{sec:eval}

\subsection{Evaluation Setup}
\noindent\textbf{Implementation:}
We implement \tbdfixed{}\footnote{\url{https://github.com/hschung1652/opencxd}} by combining MacSim~\cite{MacSim2025} with OpenSSD~\cite{openssd_acm_paper}. Parts from SkyByte's implementation of MacSim were modified and used in \tbdfixed{} as well.
For the host-side simulation, we use the latest \texttt{master} branch of MacSim extended with CXL.mem support, modeling 8 Skylake CPU cores with latency parameters.
To evaluate context switching behavior, we configure up to 3 threads per core.
On the device side, we adopt the latest OpenSSD platform, DaisyPlus~\cite{CRZTechnologyDaisyOpenSSD}, and extend its firmware with state-of-the-art CXL-SSD components described in §\ref{sec:back_sota}.
The specifications of both the DaisyPlus platform and the host system are listed in Tables~\ref{tab:eval_setup_openssd} and~\ref{tab:eval_setup_host}.

\begin{table}[!h] 
    \centering
    \vspace{-5pt}
    \caption{Specifications of the OpenSSD platform.} %\cmgt{JH: Tab2, 3 둘다 그대로 쓰시면 됩니다. 스펙 동일함.}}
    \vspace{-5pt}
    \label{tab:eval_setup_openssd}
       \scriptsize
       \begin{tabular}{|l||l|}
            \hline
            \cellcolor{gray!15}SoC & \begin{tabular}[c]{@{}l@{}}Xilinx Zynq UltraScale+ ZU17EG,\\with ARM Cortex-A53 Core\end{tabular}
            %& HYU Tiger4 SSD Controller (NVMe Controller)
            \\\hline
            \cellcolor{gray!15}NAND Module & 256GB, 4 Channel \& 8 Way
            \\\hline
            \cellcolor{gray!15}Interconnect & PCIe Gen3 $\times$ 16 End-Points
            \\\hline
            \cellcolor{gray!15}DRAM & 2GB LPDDR4 @ 2400MHz \\\hline
            %FTL & Page-based L2P Mapping Table, On-demand GC\\\hline
    \end{tabular}
    \vspace{-10pt}
\end{table}

\begin{table}[!h] 
    \centering
    \caption{Specifications of the host system.}
    \vspace{-5pt}
    \label{tab:eval_setup_host}
       \scriptsize
       \begin{tabular}{|l||l|} 
        \hline
        \cellcolor{gray!15}CPU & \begin{tabular}[c]{@{}l@{}}Intel(R) Core(TM) i7-14700K CPU\\@ 5.60GHz (28~cores)\end{tabular} \\ 
        \hline
        \cellcolor{gray!15}Memory & 32GB DDR5 \\ 
        \hline
        \cellcolor{gray!15}OS & Ubuntu 24.04.2, Linux Kernel 6.11.0 \\
        \hline
        \end{tabular}
    \vspace{-0pt}
\end{table}

\noindent\textbf{Evaluation Methodology:}
We evaluate \tbdfixed{} using memory traces from SkyByte~\cite{zhang2025skybyte}, a state-of-the-art CXL-SSD study, which includes seven workloads: \texttt{bc}, \texttt{bfs-dense}, \texttt{dlrm}, \texttt{radix}, \texttt{srad}, \texttt{tpcc}, and \texttt{ycsb}.
Each workload is executed for one million memory accesses, except for \texttt{bfs-dense}, which completes its full trace before reaching that threshold.
Note that while SkyByte is also built on MacSim, its SSD backend is implemented using the SimpleSSD~\cite{SimpleSSD2020} simulator (§\ref{sec:motiv}).
For a fair comparison, we modify SkyByte’s latency parameters to match the NAND and DRAM characteristics of our OpenSSD setup (see Table~\ref{tab:eval_setup_openssd}).
Both systems perform SSD data prefilling and host-side memory warm-up before executing the benchmarks.

\begin{figure}[!h]
    \vspace{-0pt}
    \centering
    \hfill
    \subfloat[]{%
        \includegraphics[width=.49\linewidth]{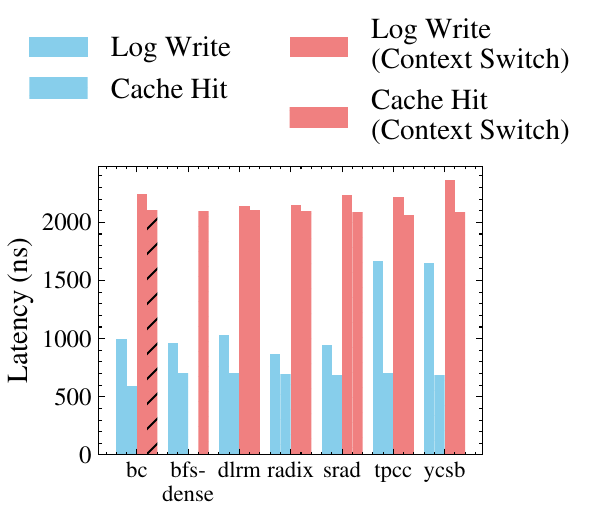}
        \vspace{-7pt}\label{subfig:eval_uops_latency}
    }
    \subfloat[]{%
        \includegraphics[width=.41\linewidth]{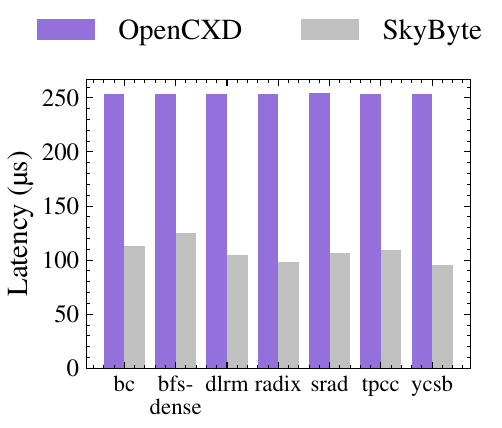}
        \vspace{-7pt}\label{subfig:eval_uops_ssdmiss_latency}
    }
    \hfill
    \vspace{-5pt}
    \caption{Average latency of key CXL-SSD optimizations, with write log inserts and DRAM cache hits as seen in \tbdfixed{} (a) and cache misses as seen in both platforms (b).}
    \label{fig:eval_uops_latency}
    \vspace{-15pt}
\end{figure}

\subsection{Re-Evaluation of State-of-the-Art CXL-SSD Optimizations}
Fig.~\ref{fig:eval_uops_latency} shows the average latencies of the performed benchmarks. A key difference between \tbdfixed{} and SkyByte can be seen in the write log insert and SSD DRAM cache hit values, as seen in Fig.~\ref{fig:eval_uops_latency}(a). 
SkyByte uses static parameters applied at compile-time to calculate these features, leading to write log insert and cache hit times to always be 640ns and 712ns respectively. 
However, \tbdfixed{} experiences differing latencies for each workload performed, and some latencies of both operations go beyond the 2$\mu s$ context switch threshold, which SkyByte uses to trigger context switches~\cite{zhang2025skybyte}. 

This shows that memory workloads with CXL-SSDs can show latency peaks well above average DRAM access times. As all I/O with CXL-SSDs involve DRAM load/store operations, they will be highly sensitive to DRAM performance penalties like latency spikes, requiring optimization to alleviate said penalties. 
Another difference seen is the SSD DRAM miss latency values in Fig.~\ref{fig:eval_uops_latency}(b). 
By accounting for NAND controller and firmware overheads, \tbdfixed{} shows 2.4$\times$ higher average latency than SkyByte across all benchmarks.
\begin{figure}[ht!]
    \vspace{-5pt}
    \centering
    \subfloat[]{%
        \includegraphics[width=.465\linewidth]{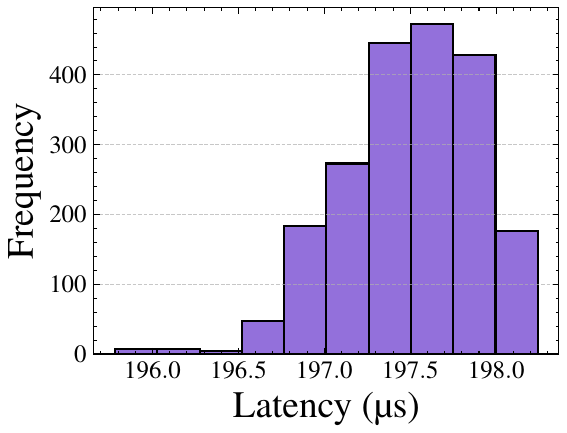}%
        \vspace{-5pt}\label{subfig:hist-a}%
    }\hfill
    \subfloat[]{%
        \includegraphics[width=.465\linewidth]{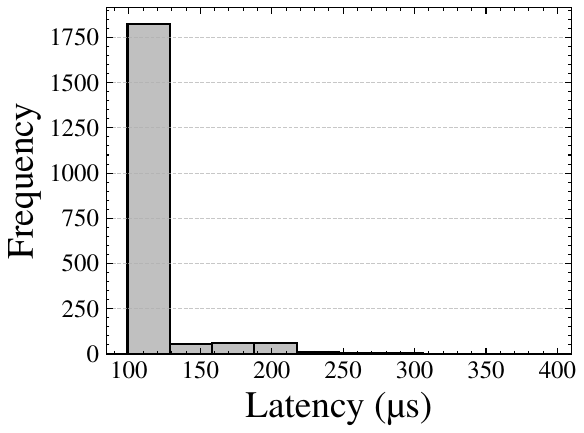}%
        \vspace{-5pt}\label{subfig:hist-b}%
    }\\
    \subfloat[]{%4
        \includegraphics[width=.465\linewidth]{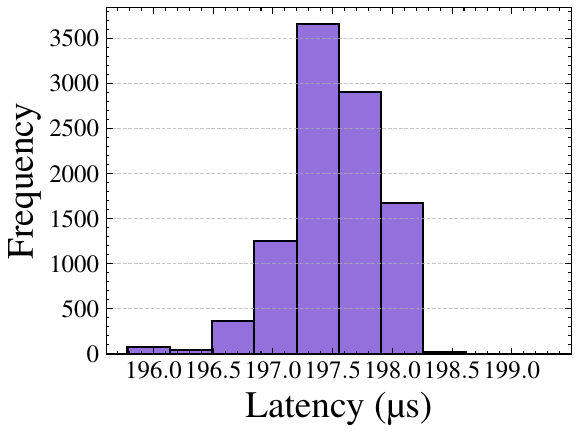}%
        \vspace{-5pt}\label{subfig:hist-c}%
    }\hfill
    \subfloat[]{%
        \includegraphics[width=.465\linewidth]{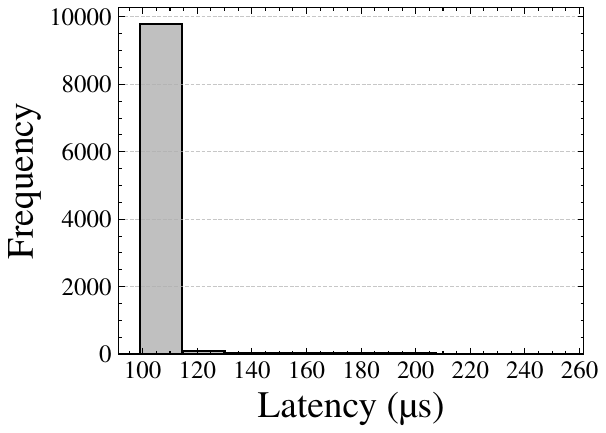}%
        \vspace{-5pt}\label{subfig:hist-d}%
    }
    \vspace{-5pt}
    \caption{Histograms of NAND read I/O latency during hit misses for \texttt{srad} ((a): \tbdfixed{}, (b): SkyByte) and \texttt{ycsb} ((c): \tbdfixed{}, (d): SkyByte).}
    \label{fig:histogram_latency}
    \vspace{-0pt}
\end{figure}

A deeper analysis of the spread of latencies can be seen in  Fig.~\ref{fig:histogram_latency}. 
The histograms show SkyByte's disproportionately high percentage of a single NAND read latency value being used, with \texttt{srad} and \texttt{ycsb} using the same value of 99.72$\mu s$ 87.2\% and 94.3\% of the time respectively. SkyByte's SSD simulation does take into account NAND I/O timeline scheduling. However, for the majority of cases, it is unable to replicate a realistic spread of latencies as seen in \tbdfixed{}. Excluding finding an applicable timeslot for the NAND I/O to take place, SkyByte  only performs mathematical calculations to apply the NAND latency \cite{skybyte_ftl}. Tracking these dynamic changes in latency is critical in evaluating CXL-SSDs, where performance is sensitive on a per memory request basis. 
Therefore, using a platform that can reflect these features like \tbdfixed{} can offer valuable insights in the design of CXL-SSD optimizations.
\begin{wrapfigure}{r}{0.24\textwidth}
    \centering
    \vspace{-3pt}
    \includegraphics[width=0.2\textwidth]{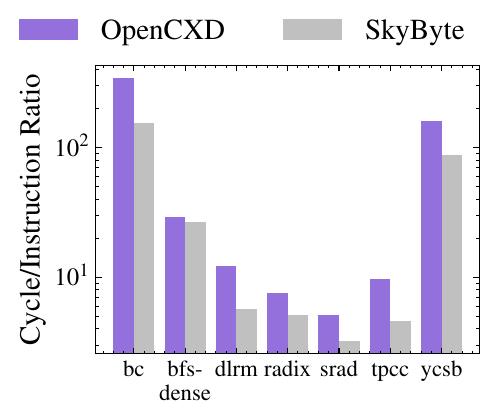}
    \vspace{-5pt}
    \caption{Log-scale comparison of CPU cycles per completed instruction for both systems.}
    \vspace{-5pt}
    \label{fig:cyc/inst_ratio}
\end{wrapfigure}

Fig~\ref{fig:cyc/inst_ratio} reveals the consequences of the previously made observations in overall performance impact. In all workloads, \tbdfixed{} required more CPU cycles to perform the amount of instructions required to perform one million memory accesses. This is in relation to the usage of context switching in an attempt to avoid the latency penalty of a NAND read. As said latencies are much higher in \tbdfixed{}, context switching over 3 threads was not enough to completely hide the read latency. These results show that more threads are required to hide the high read latency via context switching, or new optimizations are required to cover NAND I/O.

\subsection{Time breakdown of CXL-SSD optimizations}

\begin{table}[!h]
\centering
\vspace{-5pt}
\caption{Average and standard deviation of operation time related to CXL-SSD optimization. Units are in nanoseconds.}
\label{tab:stdev_optimizations}
\footnotesize
\begin{tabular}{|c|c|c|c|c|} 
\hline
\multicolumn{2}{|c|}{Workloads}                                                              & \begin{tabular}[c]{@{}c@{}}Check~DRAM~\\Cache\end{tabular} & \begin{tabular}[c]{@{}c@{}}Insert DRAM\\Cache entry\end{tabular} & \begin{tabular}[c]{@{}c@{}}Check~Write\\Log\end{tabular}  \\\hline
\hline
\multirow{2}{*}{\texttt{srad}} & Average                                                     & 37.02                                                        & 32.04                                                            & 170.86                                                      \\ 
\cline{2-5}
                      & \begin{tabular}[c]{@{}c@{}}Stddev\end{tabular} & 29.44                                                        & 29.93                                                            & 54.57                                                       \\ 
\hline
\multirow{2}{*}{\texttt{ycsb}} & Average                                                     & 36.31                                                        & 34.93                                                            & 183.2                                                       \\ 
\cline{2-5}
                      & \begin{tabular}[c]{@{}c@{}}Stddev\end{tabular} & 29.79                                                        & 29.59                                                            & 30.03                                                       \\
\hline
\end{tabular}
\vspace{-0pt}
\end{table}

Table \ref{tab:stdev_optimizations} shows a breakdown of operation overhead of state of the art CXL-SSD optimizations run on the SSD controller. While overhead for DRAM cache operation on average are seen to be around 30ns, the standard deviation is also 30ns, showing a variance in operation overhead. The overhead for checking the write log index has notably higher operation overhead, showing a clear divide of operation characteristics within the CXL-SSD controller.

\subsection{NAND Parallelism-Orientated Log Compaction}

One of the key advantages of \tbdfixed{} is its ability to reconfigure the firmware logic on the OpenSSD platform.
This flexibility allows us to experiment with CXL-SSD–specific optimizations, particularly those that exploit NAND parallelism.
As a case study, we redesigned the log compaction mechanism to leverage parallel NAND channels.
Originally, log compaction processes NAND pages sequentially: it checks the first-level index to locate modified pages, loads them into memory if not cached, merges buffered cachelines from the second-level index, and flushes the merged result to NAND (§\ref{sec:back_sota}).
In our parallel version, \tbdfixed{} first scans and tracks all required NAND pages, batches the corresponding I/O requests, and then issues them simultaneously, enabling channel-level parallelism during compaction.
\begin{wrapfigure}{br}{0.24\textwidth}
    \centering
    \vspace{-9pt}
%\begin{figure}[!h]
%    \centering
    \includegraphics[width=0.19\textwidth]{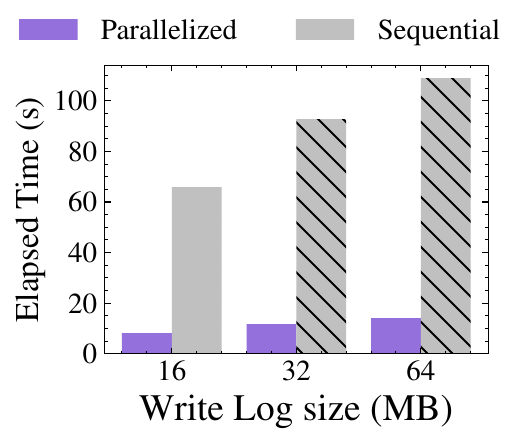}
    \vspace{-5pt}
    \caption{Comparison of write log compaction based on exploiting NAND parallelism across different write log sizes.}
    \label{fig:eval_compaction_comp}
    \vspace{-5pt}
%\end{figure}
\end{wrapfigure}
The evaluation of this optimization can be seen in Fig.~\ref{fig:eval_compaction_comp}, where improvements of up to 8$\times$ is seen across all write log sizes.
These results not only show the performance uplift of considering NAND parallelism, but also \tbdfixed{}'s ability to demonstrate implemented optimizations in real hardware.

%% file: conc.tex
%\subsection{\cb{Concurrency Support Limitations}}
\section{Conclusion}

We present \tbdfixed{}, a hybrid evaluation framework that combines cycle-accurate x86 system simulation with real SSD firmware execution on an OpenSSD platform. 
By bridging simulation and hardware, \tbdfixed{} captures critical device-level behaviors, such as DRAM latency spikes and low-level NAND controller and SSD firmware overheads, that prior simulation-only approaches overlook. 
Our results highlight the necessity of real-device-guided evaluation for accurate analysis and design of CXL-SSDs.

%\cred{yk: 1-2줄 기존 문제점을 도입으로 하고, 그래서, 제안하는 OpenCXD를 소개하지요.}

%% file: paper.bbl
\begin{thebibliography}{10}

\bibitem{kwon2018memorywall}
Y.~Kwon and M.~Rhu, ``Beyond the memory wall: a case for memory-centric hpc
  system for deep learning,'' in {\em Proceedings of the 51st Annual IEEE/ACM
  International Symposium on Microarchitecture (MICRO)}, 2018.

\bibitem{derbyshire2025memory}
K.~Derbyshire, ``{Memory Wall Problem Grows With LLMs},'' {\em Semiconductor
  Engineering}, 2025.
\newblock Last accessed: 2025-05-03.

\bibitem{zheng2023ocs}
Q.~Zheng, J.~Lee, D.~A. Manno, and G.~Grider, ``{Toward Standardized, Open
  Object-Based Computational Storage For Large-Scale Scientific Data
  Analytics},'' in {\em Proceedings of the 8th International Parallel Data
  Systems Workshop (PDSW)}, 2023.

\bibitem{gholami2024memorywall}
A.~Gholami, Z.~Yao, S.~Kim, C.~Hooper, M.~W. Mahoney, and K.~Keutzer, ``{AI and
  Memory Wall},'' {\em IEEE Micro}, vol.~44, no.~03, 2024.

\bibitem{oliveira2023extendingmemory}
G.~F. Oliveira, S.~Ghose, J.~Gómez-Luna, A.~Boroumand, A.~Savery, S.~Rao,
  S.~Qazi, G.~Grignou, R.~Thakur, E.~Shiu, and O.~Mutlu, ``{Extending Memory
  Capacity in Modern Consumer Systems With Emerging Non-Volatile Memory:
  Experimental Analysis and Characterization Using the Intel Optane SSD},''
  {\em IEEE Access}, vol.~11, 2023.

\bibitem{badam2011ssdalloc}
A.~Badam and V.~S. Pai, ``{SSDAlloc: hybrid SSD/RAM memory management made
  easy},'' in {\em Proceedings of the 8th USENIX Conference on Networked
  Systems Design and Implementation (NSDI)}, 2011.

\bibitem{jung2022cxlssd}
M.~Jung, ``{Hello bytes, bye blocks: PCIe storage meets compute express link
  for memory expansion (CXL-SSD)},'' in {\em Proceedings of the 14th ACM
  Workshop on Hot Topics in Storage and File Systems (HotStorage)}, 2022.

\bibitem{cxl31spec}
{CXL Consortium}, ``{Compute Express Link (CXL) Specification Revision 3.1}.''
  \url{https://computeexpresslink.org/wp-content/uploads/2024/02/CXL-3.1-Specification.pdf},
  2023.
\newblock Accessed: 2025-05-29.

\bibitem{pcisig_specs}
{PCI-SIG}, ``{PCI-SIG Specifications}.''
  \url{https://pcisig.com/specifications}.
\newblock Accessed: 2025-05-29.

\bibitem{das2024cxlintro}
D.~Das~Sharma, R.~Blankenship, and D.~Berger, ``{An Introduction to the Compute
  Express Link (CXL) Interconnect},'' {\em ACM Computing Surveys}, vol.~56,
  no.~11, 2024.

\bibitem{zhang2025skybyte}
H.~Zhang, Y.~Xue, Y.~E. Zhou, S.~Li, and J.~Huang, ``{SkyByte: Architecting an
  Efficient Memory-Semantic CXL-based SSD with OS and Hardware Co-design},'' in
  {\em Proceedings of the 2025 IEEE International Symposium on High Performance
  Computer Architecture (HPCA)}, 2025.

\bibitem{nvme_spec}
{NVM Express Inc.}, ``{NVM Express Specification}.''
  \url{https://nvmexpress.org/developers/nvme-specification}, 2011.
\newblock Last Accessed: 2025-05-10.

\bibitem{MacSim2025}
{Georgia Institute of Technology}, ``{MacSim: A Heterogeneous Architecture
  Timing Model Simulator}.'' \url{https://github.com/gthparch/macsim}, 2025.
\newblock Accessed: 2025-05-29.

\bibitem{gem5}
{The gem5 Community}, ``{The gem5 Simulator System}.''
  \url{https://www.gem5.org/}, 2025.
\newblock Accessed: 2025-05-29.

\bibitem{SimpleSSD2020}
{CAMELab}, ``{SimpleSSD 2.0.12 Documentation}.''
  \url{https://docs.simplessd.org/en/v2.0.12/}, 2020.
\newblock Accessed: 2025-05-29.

\bibitem{kim2009flashsim}
Y.~Kim, B.~Tauras, A.~Gupta, and B.~Urgaonkar, ``{FlashSim: A Simulator for
  NAND Flash-Based Solid-State Drives},'' in {\em Proceedings of the 2009 First
  International Conference on Advances in System Simulation (SIMUL)}, 2009.

\bibitem{li2025bytefs}
S.~Li, Y.~E. Zhou, H.~Ren, and J.~Huang, ``{ByteFS: System Support for
  (CXL-based) Memory-Semantic Solid-State Drives},'' in {\em Proceedings of the
  30th ACM International Conference on Architectural Support for Programming
  Languages and Operating Systems (ASPLOS)}, 2025.

\bibitem{zhan2024romefs}
Y.~Zhan, H.~Hu, X.~Yang, S.~Wang, Q.~Cao, H.~Jiang, and J.~Yao, ``{RomeFS: A
  CXL-SSD Aware File System Exploiting Synergy of Memory-Block Dual Paths},''
  in {\em Proceedings of the 2024 ACM Symposium on Cloud Computing (SoCC)},
  2024.

\bibitem{wang2025cxlsim}
Y.~Wang, Z.~Wang, F.~Meng, Y.~Wang, Y.~Ou, L.~Wu, W.~Hong, X.~Ge, and J.~Cao,
  ``{A Full-System Simulation Framework for CXL-Based SSD Memory System},''
  {\em arXiv preprint arXiv:2501.02524}, 2025.

\bibitem{DasSharma2022FMS}
D.~D. Sharma, ``{Transforming the Data-Centric World},'' in {\em Flash Memory
  Summit (FMS)}, Aug. 2022.
\newblock Keynote presentation.

\bibitem{MicrochipXpressConnect2020}
{Microchip Technology Inc.}, ``{XpressConnect™ PCIe® Gen 5 and CXL™
  Retimer Family}.''
  \url{https://iotdesignpro.com/sites/default/files/component_datasheet/XpressConnect-PCIe-Retimers-Datasheet.pdf},
  Nov. 2020.
\newblock Accessed: 2025-05-29.

\bibitem{simplessd_micro}
D.~Gouk, M.~Kwon, J.~Zhang, S.~Koh, W.~Choi, N.~S. Kim, M.~Kandemir, and
  M.~Jung, ``Amber: enabling precise full-system simulation with detailed
  modeling of all ssd resources,'' in {\em Proceedings of the 51st Annual
  IEEE/ACM International Symposium on Microarchitecture (MICRO)}, 2018.

\bibitem{openssd_acm_paper}
J.~Kwak, S.~Lee, K.~Park, J.~Jeong, and Y.~H. Song, ``{Cosmos+ OpenSSD: Rapid
  Prototype for Flash Storage Systems},'' {\em ACM Transactions on Storage},
  vol.~16, no.~3, 2020.

\bibitem{SKhynixCXL2025}
{SK hynix Inc.}, ``{SK hynix Completes Customer Validation of CXL 2.0-based
  DDR5}.''
  \url{https://news.skhynix.com/sk-hynix-completes-customer-validation-of-cxl-based-ddr5/},
  2025.
\newblock Accessed: 2025-05-29.

\bibitem{bandslim}
J.~Park, C.-G. Lee, S.~Hwang, S.~Yang, J.~Noh, W.~Chung, J.~Lee, and Y.~Kim,
  ``{BandSlim: A Novel Bandwidth and Space-Efficient KV-SSD with an
  Escape-from-Block Approach},'' in {\em Proceedings of the 53rd International
  Conference on Parallel Processing (ICPP)}, 2024.

\bibitem{byteexpress}
J.~Park, J.~Lee, and Y.~Kim, ``{ByteExpress: A High-Performance and
  Traffic-Efficient Inline Transfer of Small Payloads over NVMe},'' in {\em
  Proceedings of the 17th ACM Workshop on Hot Topics in Storage and File
  Systems (HotStorage)}, 2025.

\bibitem{kwon2023cxlprefetch}
M.~Kwon, S.~Lee, and M.~Jung, ``{Cache in Hand: Expander-Driven CXL Prefetcher
  for Next Generation CXL-SSD},'' in {\em Proceedings of the 15th ACM Workshop
  on Hot Topics in Storage and File Systems (HotStorage)}, 2023.

\bibitem{simplessd_pal_2024}
{SimpleSSD Project}, ``{Platform Abstraction Layer (PAL) – SimpleSSD v2.0.12
  Documentation}.'' \url{https://docs.simplessd.org/en/v2.0.12/v2.0/pal.html},
  2024.
\newblock Accessed: 2025-06-02.

\bibitem{simplessd_pal2_cc}
{SimpleSSD Project}, ``{PAL2.cc Source Code}.''
  \url{https://github.com/SimpleSSD/SimpleSSD/blob/2.0/pal/old/PAL2.cc}, 2024.
\newblock Accessed: 2025-06-02.

\bibitem{AxboeFio2022}
J.~Axboe, ``{Flexible I/O Tester (fio)}.'' \url{https://github.com/axboe/fio},
  2022.
\newblock Accessed: 2025-05-29.

\bibitem{libnvme_ioctl}
linux nvme, ``{libnvme: C Library for NVM Express on Linux}.''
  \url{https://github.com/linux-nvme/libnvme/blob/master/src/nvme/ioctl.h},
  2025.
\newblock Accessed: 2025-05-29.

\bibitem{CRZTechnologyDaisyOpenSSD}
{CRZ.TECHNOLOGY}, ``{Daisy+ OpenSSD}.''
  \url{https://www.mangoboard.com/main/view.asp?idx=1056&cate1=9}.
\newblock Accessed: 2025-05-29.

\bibitem{skybyte_ftl}
{Zhang, Haoyang and Xue, Yuqi and Zhou, Yirui Eric and Li, Shaobo and Huang,
  Jian}, ``{ftl.cc Source Code}.''
  \url{https://github.com/platformxlab/skybyte/blob/main/src/SkyByte-Sim/ftl.cc},
  2025.
\newblock Accessed: 2025-06-02.

\end{thebibliography}
